\documentclass[useAMS,usenatbib]{mn2e}

\usepackage{graphicx}
\usepackage{color}
\newcommand{\tr}{}
\title[Finding $z\sim5$ Analogues]{Establishing an analogue population for the most distant galaxies}
\author[E.~R.~Stanway \& L.~J.~M.~Davies]{Elizabeth R. Stanway$^{1}$\thanks{E-mail:
e.r.stanway@warwick.ac.uk} and Luke J.~M.~Davies$^{2,3}$\\
$^{1}$Department of Physics, University of Warwick, Gibbet Hill Road, Coventry, CV4 7AL, UK\\
$^{2}$H.~H.~Wills Physics Laboratory, University of Bristol, Tyndall Avenue, Bristol, BS8 1TL, UK\\
$^{3}$ICRAR, The University of Western Australia, 35 Stirling Highway, Crawley, WA 6009, Australia}

\begin{document}

\date{}

\pagerange{\pageref{firstpage}--\pageref{lastpage}} \pubyear{2013}

\maketitle

\label{firstpage}

\begin{abstract}
Lyman break analogues (LBAs) are local galaxies selected to match a
more distant (usually $z\sim3$) galaxy population in luminosity, UV-spectral 
slope and physical characteristics, and so provide an accessible
laboratory for exploring their properties. However, as the Lyman break
technique is extended to higher redshifts, it has become clear that
the Lyman break galaxies (LBGs) at $z\sim3$ are more massive,
luminous, redder, more extended and at higher metallicities than their
$z\sim5$ counterparts. Thus extrapolations from the existing LBA
samples (which match $z=3$ properties) have limited value for
characterising $z>5$ galaxies, or inferring properties unobservable at
high redshift.  We present a new pilot sample of twenty-one compact
star forming galaxies in the local ($0.05<z<0.25)$ Universe, which are
tuned to match the luminosities and star formation
volume densities observed in $z\ga5$ LBGs. Analysis of optical emission
line indices suggests that these sources have typical metallicities of
a few tenths Solar (again, consistent with the distant population). We
also present radio continuum observations of a subset of this sample (13 sources) and
determine that their radio fluxes are consistent with those inferred
from the ultraviolet, precluding the presence of a heavily obscured
AGN or significant dusty star formation.
\end{abstract}


\section{Introduction}
Lyman break analogues (LBAs) are a class of compact, UV-luminous
galaxies in the local universe that are selected to match the more
distant Lyman break galaxy (LBG) population in luminosity, UV-spectral
slope and physical characteristics. Lyman break galaxies are named for
their dominant spectral feature - a strong rest-frame ultraviolet
continuum, which demonstrates a significant flux decrement at the
wavelengths corresponding to 912\AA\ and 1216\AA\ in the galaxy
rest-frame.  Below 912\AA\ photons are sufficiently energetic to
ionise neutral Hydrogen in the intergalactic medium and so are
efficiently scattered from the line of sight before reaching the
observer. Between 912\AA\ and 1216\AA\ the same intergalactic medium
can also scatter light in narrow absorption lines representing
electron transitions in the Lyman series of Hydrogen (particularly
Lyman-$\alpha$). Given that the density of gas clouds in the
intergalactic medium rises sharply with increasing redshift, and each
cloud absorbs at wavelengths corresponding to its own redshift, the
resulting `forest' of redshifted lines can lead to a significant
amount of flux lost from the line of sight.

At $z\sim3$, the cumulative effect of scattering in Lyman-$\alpha$
forest leads to a sharp break in the spectrum, such that about half
the flux shortwards of 1216\AA\ is lost and below the second break at
912\AA\ reduces to negligible levels. Thus a source will appear absent
in a photometric image taken in a filter lying below this wavelength
and is said to have ``dropped-out''. Therefore $z\sim3$ Lyman break 
galaxies, identified by their broadband imaging, are termed ``U-dropouts''
or sometimes ``U-drops''. At higher redshifts still, the Lyman-$\alpha$
forest is denser, with the decrement across 1216\AA\ in the rest frame 
exceeding 90\%, and the break has moved to higher observed wavelengths. Thus
$R-$drops can be selected at $z\sim5$, $I-$drops at $z\sim6$ and so on.

While rest-frame ultraviolet-selected LBGs are
our primary source for understanding galaxy formation and the `normal'
galaxy population in the distant Universe, our knowledge of them is
necessarily limited by their faint apparent magnitudes, small
projected sizes, and extreme redshifts.  The interpretation of these
galaxies can be greatly enhanced by the consideration of local
analogues for which more extensive and more detailed data can be
obtained.  The study of local galaxies as potential LBAs has developed
rapidly over recent years. Heckman and collaborators have built on
their early work \citep{2005ApJ...619L..35H,2007ApJS..173..441H}, identifying a sample of
ultra-compact, ultraviolet-luminous galaxies that are a close mirror
to $z\sim3$ LBGs. Such sources are selected in data from the Galaxy
Evolution Explorer (GALEX) survey and subsequent studies have explored
their properties in infrared photometry \citep[e.g.][]{2011A&A...536L...7I},
X-rays \citep{2011ApJ...731...55J}, ultraviolet spectroscopy \citep{2011ApJ...730....5H}
and morphology \citep[e.g.][]{2011ApJ...726L...7O}.

However, work by \citet{2012ApJ...745...96H} and others has shown the
importance of consistency in selection techniques across
redshift. Recent colour selected samples of $z\sim2$ candidates such as the
BX/BM samples have
been described as `Lyman break galaxies' but are in fact selected
based on rest-frame optical, rather than ultraviolet colours
\citep{2004ApJ...604..534S}.  As \citet{2012ApJ...745...96H} found,
this difference in selection technique, while identifying galaxies at
the predicted redshift, biases galaxy samples towards a quantitatively
different population. Similarly $z=0.2$ `Green Peas' \citep[star
  forming galaxies selected for their strong rest-frame optical
  emission lines and compact morphology,][]{2009MNRAS.399.1191C} are
distinct from analogues selected from rest-UV photometry at the same
redshift \citep[e.g.][]{2011ApJ...730....5H}. While all the galaxies
identified by these techniques are actively star-forming, their
derived properties such as star formation rate, dust extinction,
stellar population age, physical size and hence physical conditions
can be very different. Given that each of these populations is deemed,
in certain ways, characteristic of its redshift, it becomes
increasingly difficult to interpret comparisons between these
discrepant populations as indicative of evolution either in the
building blocks of today's massive galaxies or of wider
volume-averaged cosmological properties.  It is vital that the
criteria used to identify galaxy populations to be compared at
different redshifts are as close as possible to being identical.

As the Lyman break technique is extended to ever higher redshifts
\citep[e.g.][]{2010MNRAS.409.1155D,2009MNRAS.400..561D,2012ApJ...754...83B,2007ApJ...670..928B},
it has become increasingly clear that the LBA population identified by
Heckman and co-workers, and the `Green Pea' selection of Cardamone et
al, provides a poor match for the properties of high redshift ($z>5$)
UV-selected galaxies. Selected to match the typical properties of
$z\sim3$ LBGs, they are too massive and often too red to provide a
good comparison to existing $5<z<8$ samples, which are less luminous
\citep[L$*_{z=6}\sim0.3$L$*_{z=3}$, e.g.][]{2007ApJ...670..928B},
younger and less massive \citep[$\sim30$\,Myr,
  M$_\mathrm{stellar}\sim$few$\times10^9$\,M$_\odot$, e.g.][although
  c.f.~Oesch et al 2013]{2007MNRAS.377.1024V} and at lower
metallicities \citep[Z$\sim0.2-0.5$Z$_\odot$,
  e.g.][]{2010MNRAS.409.1155D} than their $z\sim3$
counterparts. Curiously, galaxies that may act as plausible analogues
for exceptionally low-metallicity star-forming galaxies (touted as
being appropriate for the very earliest galaxies at $z>9$, but yet to
be observationally confirmed) have received more attention
\citep[e.g.][]{2011A&A...536L...7I} than those at a few tenths of
solar, as appropriate to $z\sim5-7$. LBGs at $z>5$ are also
significantly more compact than their lower redshift counterparts
\citep{2010ApJ...709L..21O}, driving the star formation density (both
in terms of volume and per unit stellar mass) to levels not seen in
the equivalent $z=3$ population.  This higher star formation density
is a key observational driver for searching for better
analogues. Higher UV-photon density changes the physical conditions
within the galaxies. Dust temperatures are likely to be higher, the
intergalactic medium warmer, and hence the collapse of molecular
clouds into stars, the ionization of the intergalactic medium and
potentially the mode of star formation itself rather different.  Thus
analogues with lower star formation rate density (as typical for the
$z=3$ population) simply {\it cannot} provide good physical models
with which to understand galaxies in the very distant universe.  If
LBAs are to inform our understanding at $z\geq5$, or extrapolate from
the UV-luminous component, then a specialized $z\sim5$-equivalent LBA
population must be established.

Through a combination of archival work and proprietary observations,
we have developed a sample of spectroscopically-confirmed,
star-forming, $z=0.05-0.20$ galaxies identified in the UV and optical
(using GALEX DR6 and the SDSS DR7), that we are exploring in detail as
analogues for the $z>5$ population. In this paper we discuss the
properties of a plausible candidate Lyman Break analogue galaxy sample, tuned
to match the physical properties of the observed $z\geq5$ galaxy
population. In section \ref{sec:cand} we identify the sample under
discussion, exploring their properties in the SDSS optical survey in
section \ref{sec:sdss}. In section \ref{sec:radio} we present new
radio imaging of a subsample of objects and discuss the results,
before extending the discussion to the implications of the sample more
generally in section \ref{sec:conc}.

Throughout, optical magnitudes are presented in the AB system. Where
necessary, we use a standard $\Lambda$CDM cosmology with
$H_0=$70\,km\,s$^{-1}$\,Mpc$^{-1}$, $\Omega_M=0.3$ and
$\Omega_\Lambda=0.7$.

\section{Candidate Selection}\label{sec:cand}

\subsection{Motivation}

While it has become clear that the intrinsic properties of the Lyman
break galaxy population evolves with redshift, it has not been clear
whether the Lyman break analogue candidate samples already suggested
or investigated is sufficiently broad to encompass a subset of good $z>5$
analogue systems. 

We explored this question using the largest, best developed LBA
sample: that identified by \citet{2005ApJ...619L..35H} and expanded in 
\citet{2007ApJS..173..441H}\footnote{While the Hoopes et al targets were
primarily identified as ultraviolet-luminous galaxies (UVLGs), they
were nonetheless motivated as an LBA sample.} and related work. These sources were selected based on a
combination of UV luminosity and UV-optical colour to mirror the
properties of the $z=3$ Lyman break galaxy population, with the
primary selection based on UV flux (and hence inferred star formation
rate) exceeding $\sim$0.3\,$L_\odot$ for $z=3$ LBGs \citep{2007ApJS..173..441H}. 
The galaxies occupy a redshift range of $0.08<z<0.30$, with a
mean of $z=0.20$. No constraint was placed on the angular size of the
systems. However, projected size (after deconvolution with the seeing) was
available from SDSS data.

\begin{figure}
\includegraphics[width=\columnwidth]{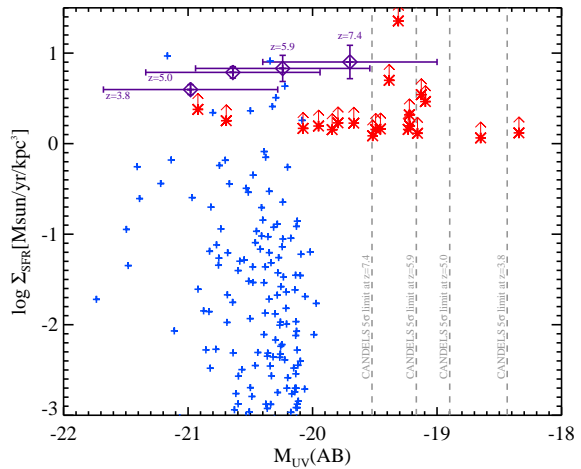}
\caption{The star formation density as function of luminosity for LBG
  samples at high redshift (diamonds), the LBA sample of Hoopes et al
  (2007, crosses) and the sample discussed in this work
  (asterisks). The current sample are shown as lower limits since all
  are essentially unresolved in SDSS data. We calculate the inferred
  star formation density from the observed 1500\AA\ flux and SDSS
  morphology, assuming for simplicity that the observed star formation
  is concentrated within a cube with a scale length given by the
  observed half-light radius of the galaxy (note that a spherical
  assumption would change each value by a factor of 4$\pi$). For LBG
  samples, we use the typical absolute magnitudes determined by
  Bouwens et al (2007), and typical sizes from Oesch et al (2010) from
  deep, space-based dropout samples. The horizontal ranges on these
  points indicate a factor of 2 in luminosity, either side of the
  typical luminosity M$^\ast$. We also show the limits in absolute
  magnitude corresponding to the `Deep' CANDELS survey target depths
  of 27.5\,AB. The standard conversion
  factor of Madau et al 1998
  ($L_{UV}=8.0\times10^{27}$\,(SFR\,/\,M$_\odot$/yr)\,ergs\,s$^{-1}$\,Hz$^{-1}$,
  using a Salpeter IMF) is assumed.\label{fig:lum-size}}
\end{figure}

One of the primary drivers for identifying analogues to distant
sources is to explore the effects of very high volume-averaged star
formation densities, as discussed above.  In figure
\ref{fig:lum-size}, we calculate the inferred star formation density
(assuming that the observed star formation is concentrated within the
observed half-light radius of the galaxy) for the catalogue of 
\citet{2007ApJS..173..441H} and compare it to the derived typical values for Lyman
break galaxy samples at high redshift, using the Schecter parameter
M$^\ast$ \citep[from][]{2007ApJ...670..928B} as a typical luminosity and the
mean physical half light radius at each redshift from \citet{2010ApJ...709L..21O}. 
Star formation rates are calculated from the rest-frame ultraviolet flux
density, using the conversion factor of \citet{1998ApJ...498..106M}, 
 $L_{UV}=8.0\times10^{27}$\,(SFR\,/\,M$_\odot$/yr)\,ergs\,s$^{-1}$\,Hz$^{-1}$,
  assuming a Salpeter IMF. As is often the case for work on Lyman break 
galaxies at $z\ga5$, no correction is made for dust extinction of
the ultraviolet flux (which is essentially unknown at $z>5$) or any
change in the IMF with redshift. Given that dust extinction at
$z\sim3$ can be as much as factor of a few, the inferred star formation
rates are considered to be lower limits, and the star formation densities
may be still higher than shown - however observations in the thermal far-infared
and submillimeter would be required to reliably determine a total star formation rate
- challenging even for ALMA at $z\sim5$. The rationale here is to compare like for
like - observed ultraviolet luminosity for observed ultraviolet luminosity - and
investigate dust properties when a sample for investigation has been defined.
As the figure clearly demonstrates, very few of the Hoopes et
al analogue sample ($\sim$3\%) have a star formation density
comparable to that observed in distant Lyman break galaxies - a
discrepancy that only becomes exacerbated as the Lyman break samples
are pushed to ever higher redshifts.

Figure \ref{fig:lum-size} also illustrates a second concern with the
use of previously identified local, luminous starbursts as Lyman break analogues: the
increasingly low luminosities being probed by deep and distant
surveys. The typical luminosity of Lyman break galaxies is firmly
established to evolve with increasing redshift, being some 1.5
magnitudes fainter at $z\sim7$ than at $z\sim3$ \citep[e.g.][]{2011MNRAS.417..717W}. Even setting aside
evolution in the physical size distribution of these sources, it
is clear that finding analogues for $z\geq5$ galaxies identified in
surveys such as CANDELS \citep[][which is probing 120\,arcmin$^{2}$ to a
uniform depth of $\sim$27.5 in ten broad wavebands]{2011ApJS..197...35G} will require the
study of less luminous and more compact galaxies than
previously considered.

\subsection{Selection criteria}

We have explored the public data releases from the Sloan Digital Sky
Survey \citep[DR7, ][]{2009ApJS..182..543A} in the optical ($ugriz$ bands), and GALEX (GR6\footnote{Available at http://galex.stsci.edu/GR6/}) in its two ultraviolet bands (FUV and NUV, centred at
1538\AA\ and 2315\AA\ respectively) for systems that might satisfy
this requirement. In order to tune our sample to better match the high
redshift population, we apply several selection criteria:

\textit{Colour:} Where \citet{2005ApJ...619L..35H} use the fairly
liberal UV-to-optical criterion of $FUV-r<2$, suitable for identifying
continuous star formation over timescales of several hundred years, we
do not rely on optical colours but rather apply a tight constraint in
the UV - analogous to a high redshift LBG selection. Beyond $z=5$, the
rest-frame optical is shifted well into the infrared and difficult to
observe. Hence candidate selection is performed through a combination
of a strong spectral break at 1216\AA\ \citep[due to absorption in the
  intervening intergalactic medium,
  e.g.][]{2007ApJ...670..928B,2004MNRAS.347L...7B,2003MNRAS.342..439S}
and a flat or relatively blue spectral slope observed as a small
photometric colour between bands in the rest-frame ultraviolet
\citep[][]{2013MNRAS.430.2885W,2005MNRAS.359.1184S}.

\begin{figure}
\includegraphics[width=\columnwidth]{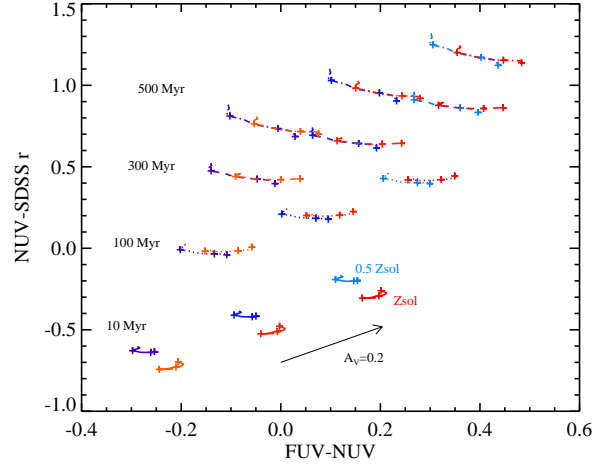}
\caption{Ultraviolet-optical colours expected of young starburst
  stellar populations at $0<z<0.3$ (increments of $\Delta z=0.1$ are
  marked with crosses on each evolutionary track). We use 
  \citet{2005MNRAS.362..799M} stellar population synthesis models for a declining initial
  starburst to determine the expected colour at four ages, and two
  metallicities and also show the effects of dust reddening with an
  A$_V$=0.2, 0.4, assuming a \citet{2000ApJ...533..682C} extinction
  law. Increasing stellar population age primarily effects the
  UV-optical colour, while the effects of dust are larger in the
  ultraviolet. A selection criterion of $FUV-NUV<0.5$ encompasses the
  bulk of young stellar populations and reasonable dust extinctions in
  the local Universe. \label{fig:col-col}}
\end{figure}

To mirror this, we require that the rest-UV slope is very close to
flat (i.e. $FUV-NUV<0.5$), targeting galaxies with young starbursts of
$<$200\,Myr (see figure \ref{fig:col-col}). We must, of course, allow
for some range of dust extinction in the population.
\citet{2007MNRAS.377.1024V} found that the spectral energy
distribution of $z\sim5$ galaxies are best fit with $A_V\sim0.3$\,mag,
a little lower than that measured at $z\sim3$ ($A_V\sim0.5$\,mag,
Shapley et al, 2003). Evidence from the rest-frame ultraviolet slopes
of candidate samples suggest Lyman break galaxies at higher redshift
may have still lower extinctions, but with considerable uncertainties
\citep[see][for a detailed discussion]{2013MNRAS.430.2885W}. To
accommodate this variation, we do not explicitly constrain the $NUV-r$
band colour, allowing for variations in stellar population age within
the sample, and moderate dust extinction. However we note that no
plausible candidates had colours redder than $NUV-r=2.2$.

\textit{Luminosity:} We tune our selection
window to FUV absolute magnitudes which are equivalent to those in
existing $z>5$ LBG samples. This requires that we probe 
galaxies with L$_{UV}$=$0.1-5$\,L$*_{z=6}$, where M$^*_{UV}$=-20.24 at $z=6$
\citep{2007ApJ...670..928B}. 

\textit{Size:} We also require that this star formation occurs in a
compact region \citep[distant galaxies have a projected half-light radius
$<$ 2\,kpc,][]{2010ApJ...709L..21O}, so as to ensure a similar star
formation volume density in the local analogue systems to that observed 
in the high redshift sources. Given the typical seeing of SDSS images,
deconvolution of very compact sources to determine their physical size
is problematic. At $z>0.2$, a physical size of 2\,kpc is identified as
only a small perturbation on the light profile due to average
seeing. Thus the requirement that the SDSS pipeline is able to
constrain the size effectively provides an upper limit on our redshift
distribution. Given the uncertainties on deconvolution of these
objects, we apply the reasonably liberal criteria that the best fit
deconvolution corresponds to a physical scale $r<3.5$ kpc, and that
the SDSS $g$-band petrosian radius is $<1.2\arcsec$. While this may
admit a fraction of sources more extended than those at high redshift,
as figure \ref{fig:lum-size} shows these lower limits select a
population with star formation densities comparable to those observed
in the distant Universe.  Note that we must constrain the physical
size in the optical, since GALEX does not have sufficient resolution
in the UV.

\textit{Additional Constraints:} In order to identify a useful pilot
sample for further study we require that our candidates have SDSS
spectroscopy, and that this spectroscopy identifies their UV
luminosity as arising from star formation rather than AGN \citep[the AGN
fraction in distant galaxy samples is known to be very small,][]{2007MNRAS.376.1393D,2005MNRAS.360L..39N}.  In the interests of further, ground-based
follow-up, we also place a declination constraint (dec\,$<-8.5\deg$)
to optimize availability to southern telescopes (e.g. ATCA, VLT, ALMA
etc.). We eliminate catalogue sources with obvious problems with
their photometry (e.g. small clumps which form part of a much larger, more
extended system, or objects whose photometry is affected by near
neighbours - a particular problem for these faint targets in the GALEX data). 

\begin{figure}
\includegraphics[width=\columnwidth]{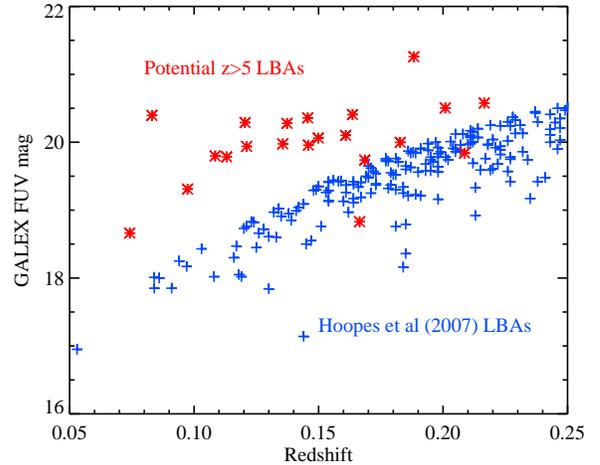}
\caption{The redshift-apparent magnitude distribution of our targets, compared to the ultraviolet-luminous galaxy sample of Hoopes et al (2007). While there is some overlap, our targets are typically fainter, and at a slightly lower mean redshift.  \label{fig:zfuv}}
\end{figure}

These criteria identify a sample of just 21 sources to investigate in
detail, as a subset of the larger, unrestricted declination, population. As figure 1 (and
also section~\ref{sec:sdss}) demonstrates, these occupy a region of parameter space distinct
from that considered by previous studies - both in luminosity and in star formation density.
Unsurprisingly these are both fainter in apparent magnitude, and at a slightly lower typical
redshift than those of the \citet{2007ApJS..173..441H} sample (see figure \ref{fig:zfuv}).

\section{Properties in the GALEX/SDSS data}\label{sec:sdss}

 \subsection{UV slope and optical morphology}

\begin{figure}
\includegraphics[width=\columnwidth]{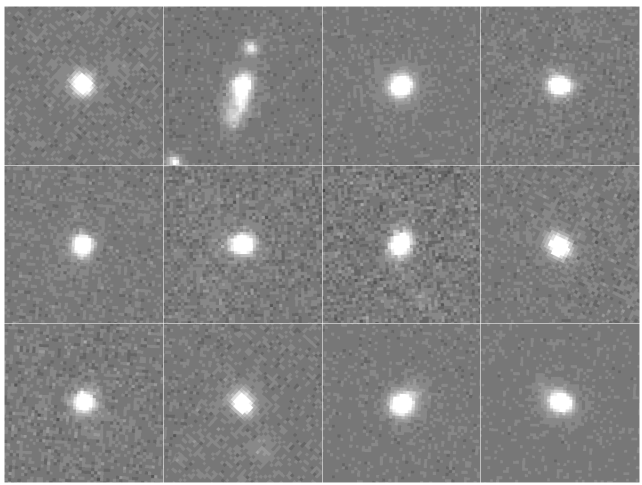}
\caption{Examples of the essentially unresolved optical morphology (SDSS $g$ band) of these candidates. Boxes are 10$\arcsec$ on a side. A few objects (for example, the second here) show more extended structure around a central core, and these may prove not to be ideal high redshift analogues, but they are included in the sample so as to explore the range of galaxies identified in catalogue data. Deconvolution suggests a typical half-light radius of $<$0.5\,arcsec.\label{fig:morph}}
\end{figure}

 The selection criteria discussed in section \ref{sec:cand}
 essentially select sources which are unresolved in the optical, as
 figure \ref{fig:morph} shows. However a few objects are selected for
 which the properties of a dominant central core satisfy our selection
 criteria, but which consist of multiple compact components or possess
 a slightly more extended, low surface brightness tail. Given that
 multiple components are often seen in $z\sim5$ samples \citep[e.g.][]{2009MNRAS.400..561D}, and that low surface brightness emission may well be
 undetectable at high redshifts, we retain these sources in our
 sample, in order to explore the full properties of this
 colour-selected population. Further observations (using either adaptive optics
 or space-based data) will be required to fully constrain the morphology of these
 sources. 

 Each source in our sample has a full suite of UV-optical photometry.  We use
 GALEX photometry (at 1300 and 2400\AA) and redshift information from
 SDSS spectroscopy, to explore the observed rest-frame ultraviolet spectral
 slope of our sample. The observed spectral slope is 
 primarily dependent on a combination of stellar population age and dust
 extinction, with the bluest colours ($f_\lambda\propto\lambda^\beta$,
 $\beta<-2$) only possible with zero dust and very low
 metallicities (see figure \ref{fig:col-col}, and also Wilkins et al
 2013 for a fuller discussion). In common with the high redshift samples
 with which we ultimately wish to compare, we are unable to straightforwardly disentangle
 the effects of dust and stellar population from a single colour, but instead consider
 the observed range of values, with a full analysis of dust deferred until more data is
 available (see discussion later). Even without separating dust and stellar population
 age, their combined effect is an important indicator of
 the properties of a galaxy, and has drawn recent attention.

Some authors
\citep[e.g. ][]{2013arXiv1306.2950B,2012ApJ...754...83B,2010ApJ...724.1524O}
have suggested that galaxies at the highest redshifts ($z>7$) are
systematically bluer than those observed at later times and can have
extreme spectral slopes, with $\beta<-3$.  Such a blue slope is
difficult to produce using normal stellar population models and may
imply the presence of very low metallicity or Population III
stars. However this observation is subject to significant
uncertainties and the colours are likely less extreme than originally
suggested \citep[see][ for detailed
  discussion]{2013MNRAS.430.2885W}. \citet{2012ApJ...756..164F},
considering the same data as \citet{2012ApJ...754...83B}, suggested
that the spectral slope, while evolving to bluer colours between
$z\sim4$ and $z\sim7$, does not require such exotic stellar
populations. By contrast, \citet{2013MNRAS.432.3520D} have suggested
that there's no strong evolution in the slope beyond $z\sim4$ and that
while the colours are typically blue ($\beta\sim-2$), they can be
straightforwardly explained with low dust, or slightly sub-solar
metallicity, populations.  Interestingly, some studies at
\tr{intermediate} redshift ($z<4$) have found \tr{relatively} little
evidence for a systematic and linear colour evolution with
redshift. \tr{Studying stacked mid-infrared (Herschel) data for
  ultraviolet-selected galaxies in photometric redshift bins at
  $z\sim1.5, 3$ and 4, \citet{2014MNRAS.437.1268H} identified a strong
  evolution in their dust extinction with stellar mass, but negligible
  evolution with redshift.  If UV spectral slope is interpreted as
  varying due purely to dust (as opposed to stellar population age, or
  metallicity, then this might suggest that somewhere around
  $z\sim4-5$ (i.e. when the first galaxies are about 1\,Gyr old), the
  colour evolution stabilises as older stellar populations (and the
  dust they generate) begin to contribute significantly to the
  observed colours.}

\begin{figure}
\includegraphics[width=\columnwidth]{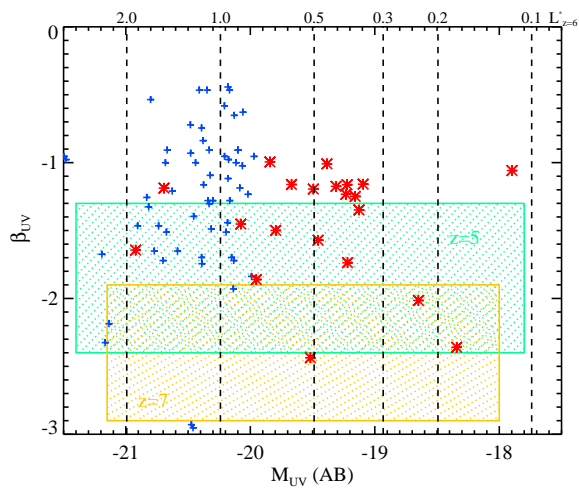}
\caption{A comparison of the $z\sim3$ LBA population (crosses), with the sample discussed here (shown as asterisks). The sample presented here forms a distinct and separate galaxy population, probing the lower luminosities, higher star formation densities and sometimes bluer rest-UV colours typical of the high-z population. Shaded regions indicate the range of parameter space that has been estimated by varied authors for LBG samples at $z=5$ and $z=7$ \citep[see][]{2013MNRAS.430.2885W}.\label{fig:lum}}
\end{figure}

\tr{Given that studies such as \citet{2014MNRAS.437.1268H} and \citet[][ who studied individually Herschel-detected $z<2$ examples]{2011ApJ...734L..12B} suggest a strong luminosity dependence on inferred dust extinction at moderate redshift, while at the highest redshifts \citet[][]{2013arXiv1306.2950B,2012ApJ...754...83B} and others have suggested a similar luminosity dependence, it is informative to examine our sample for any similar trend.} 
In figure \ref{fig:lum} we plot the rest-frame ultraviolet spectral
slope for our sample, as a function of absolute magnitude. As can be
seen, the sample probes slopes in the range $-2.5<\beta<-1.0$,
comparable to the typical slopes seen at $z\sim5-7$ \citep{2011MNRAS.417..717W,2013MNRAS.430.2885W, 2010MNRAS.409.1155D,2005MNRAS.359.1184S},
 and shows a similar trend
towards bluer slopes at lower luminosities to that suggested in the distant
universe.  This trend is, admittedly, rather weak in our small pilot
sample. Nonetheless, we note that six of our sources (29\%) have
spectral slopes with $\beta<-1.5$, substantially steeper than that
seen in the $z=3$ population and its analogues \citep[$\beta\sim-0.9$,][]{2003ApJ...588...65S}.

 \subsection{Emission Line indices}
\begin{figure}
\includegraphics[width=\columnwidth]{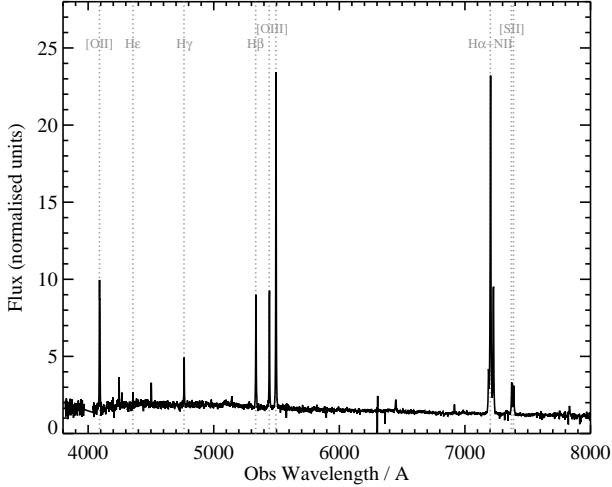}
\caption{An example of the emission line spectrum typical of these galaxies. Several prominent emission lines related to star formation and metallicity fall into the observed frame optical spectrum. This example lies at $z=0.097$.\label{fig:spec}}
\end{figure}

All the targets in this sample were identified as star-forming
galaxies on the basis of their SDSS spectra, and show the typical
strong and narrow optical emission lines (see figure \ref{fig:spec}).
In order to evaluate the effectiveness of this sample as LBAs, an
important property that must be considered is the metallicity of the
dominant stellar population. This is still somewhat poorly constrained
at higher redshifts where optical spectroscopy is beyond the grasp of
current instruments. As mentioned in the previous section, extremely
low metallicities ($<0.001$\,Z$_\odot$) have been proposed for the most
distant sources, although it is not clear these are strictly
necessary.

At $z\sim5-6$ the metallicity of typical Lyman break galaxies
is slightly better known. Fitting of the full spectral energy
distribution of photometrically selected galaxies suggest that
reasonable fits can be obtained using synthetic stellar populations
with metallicities of a few tenths solar \citep[e.g.][]{2007MNRAS.377.1024V}, while solar metallicity models tend to suggest
implausibly old stellar populations at these early times. Similarly,
the blue spectral slopes at high redshift, if interpreted as a
metallicity indicator yield approximate values of
$\sim$0.25\,Z$_\odot$ \citep{2010MNRAS.409.1155D}. This is consistent with the observed metallicity
of absorption line systems in the interstellar medium which is seen to 
decrease by $\sim1$ dex between $z=0$ and $z=5$ \citep[and perhaps more rapidly
at still higher redshifts, e.g.][]{2012ApJ...755...89R}. Similarly, Gamma
Ray Burst host galaxies at $z>4$ \citep[which are believed to be very low mass, ][, but must be star-forming in order
to generate the short-lived GRB progenitors]{2012ApJ...754...46T}, have been found to have
typical metallicities $<0.15$\,Z$_\odot$ \citep[see compilation in ][]{2013MNRAS.428.3590T}. 

Thus we would expect good analogues for the distant galaxy population
to have metallicities of a few tenths of solar, perhaps decreasing as
we search for analogues of galaxies at either higher redshifts or
lower luminosities. This is, of course, a crude estimate, and will
depend somewhat on the method and species used to make the
measurement. We might expect good analogues (which should be young
starbursts) to show an enhancement in $\alpha$-process elements, for
example \citep[e.g.][]{2008ApJ...681.1183K}.

\begin{figure}
\includegraphics[width=\columnwidth]{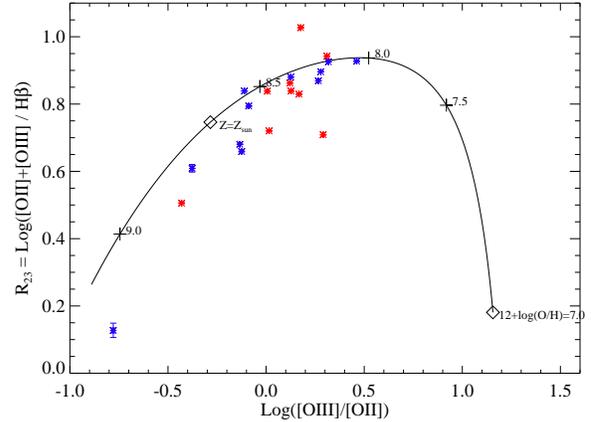}
\caption{The variation in optical emission line ratios, established as metallicity indicators, in our sample. The solid line shows the mean metallicity as a function of emission line ratio strength in the SDSS galaxy population, calibrated by Maiolino et al 2008 - note that the SDSS galaxy population shows considerable scatter around this mean relation. We adopt a Solar metallicity of 12+log(O/H)=8.69. Galaxies with radio data (see section \ref{sec:radio}) are shown in red, the remainder of the sample in blue. Formal errors on the measured equivalent widths (propagated to the line ratios) are too small to plot (with a few exceptions), however we note that uncertainty in the continuum for these relatively faint sources may introduce errors at the $<10$\% (0.05 dex) level. The sources in this sample probe moderate metallicities, with a few outliers.\label{fig:metallicity1}}
\end{figure}

\begin{figure}
\includegraphics[width=\columnwidth]{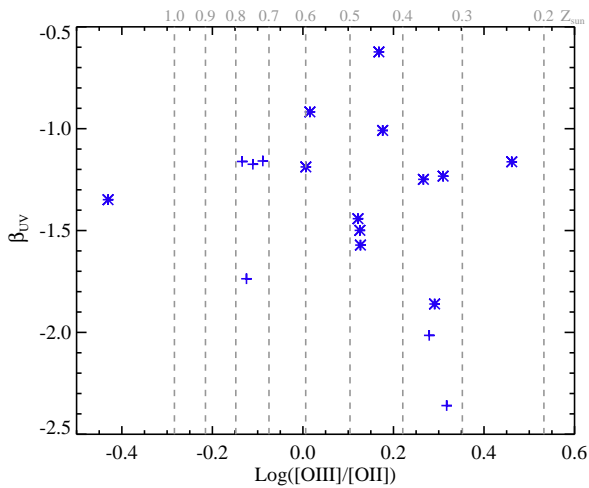}
\caption{The variation in rest-UV spectral slope $\beta$ with metallicity (indicated here by the ratio of oxygen emission line strengths) based on the calibration of Maiolino et al 2008 in figure \ref{fig:metallicity1}. The bulk of this sample probes $0.2<$\,Z$_\odot<0.6$, exploring the same metallicity range inferred for the $z\sim5$ population. There is tentative evidence for bluer spectral slopes (or at least more dispersion in the spectral slope) with decreasing metallicity.\label{fig:zbeta}}
\end{figure}

In figure \ref{fig:metallicity1}, we plot the classic metallicity
indicators $R_{23}$=([OII]+[OIII])/H$_\beta$ and log([OIII]/[OII]) for
our sample. Line fluxes were extracted from the SDSS spectroscopic
catalog for our candidate sources, and tested for consistency against
the OSSY determinations by
\citet{2011ApJS..195...13O}\footnote{http://gem.yonsei.ac.kr/$\sim$ksoh/wordpress/?page\_id=18}. Overplotted
on figure \ref{fig:metallicity1}, we show the calibrated empirical
relation between these quantities and metallicity from
\citet{2008A&A...488..463M}. As can be seen, the use of the oxygen
ratio can break the well-known metallicity degeneracy in the $R_{23}$
index, placing all of our candidate sample on the upper arm of the
metallicity relation. The bulk of our sample (17/21 objects) shows
metallicities in the range $8.00<$12+log(O/H)$<8.69$
(i.e. $0.2<$\,Z$_\odot<1.0$, a few tenths Solar). A few outliers in the sample either have
higher metallicities or poorly constrained metallicity indices (due to
redshifting of one or more emission lines into regions of low signal
to noise in the spectroscopy). We note that the blue colours targeted
in our sample are clearly possible even at super-Solar metallicities,
and will consider the effects of this higher metallicity in individual
objects during later analysis. However, as figure \ref{fig:zbeta}
demonstrates, at least half (11/21 objects) of our sample have
measured optical spectral indices consistent with the metallicity
range $0.2<$Z$_\odot<0.5$ that may be appropriate for high redshift
analogues.

Evolution in the rest-frame ultraviolet spectra of distant sources is
expected to be driven, at least in part, by evolution in the cosmic
mean metallicity at early times. This has been observed between $z=5$
and $z=3$ (e.g. Douglas et al 2010), and inferred photometrically at
higher redshifts (although with some uncertainty, see Wilkins et al
2013). Figure \ref{fig:zbeta} suggests an intriguing trend is
present in our sample.  While the number statistics are small, there
appears to be a trend towards bluer rest-frame ultraviolet spectral
slopes with decreasing metallicity within our sample. Certainly there
is a larger dispersion in the spectral slopes for sources at
Z$<0.5$Z$_\odot$ than for sources around Solar metallicity. However we
caution that this sample is too small to disentangle the effects of
trends of luminosity (see figure \ref{fig:lum}) and stellar population
age from a pure metallicity evolution. Thus we do not quantify the
trend here, due to the small sample statistics, but will investigate
this further in future work.

 \subsection{Optical Colour}
In addition to the UV-selected LBA samples discussed above, a second category of local object has been suggested as Lyman break analogues: the optically-selected Green Pea population \citep{2009MNRAS.399.1191C}. These sources are identified from their broad-band photometry as having an excess of flux in the $r$-band, interpreted as indicative of strong [OIII] emission at $0.112<z<0.360$. Hence this selection is designed to identify strongly star-forming galaxies in the local universe.

\begin{figure}
  \includegraphics[width=0.495\columnwidth]{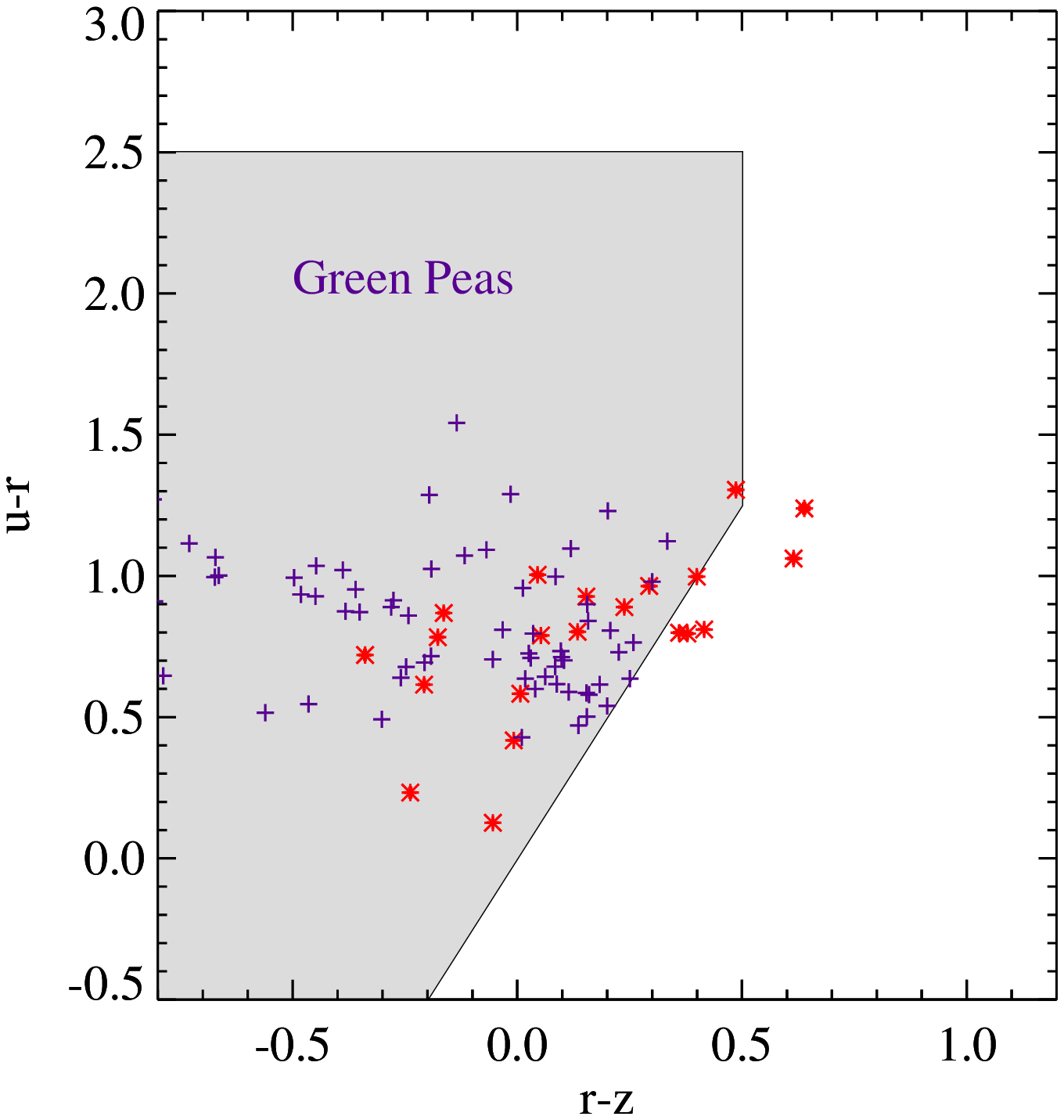}
  \includegraphics[width=0.495\columnwidth]{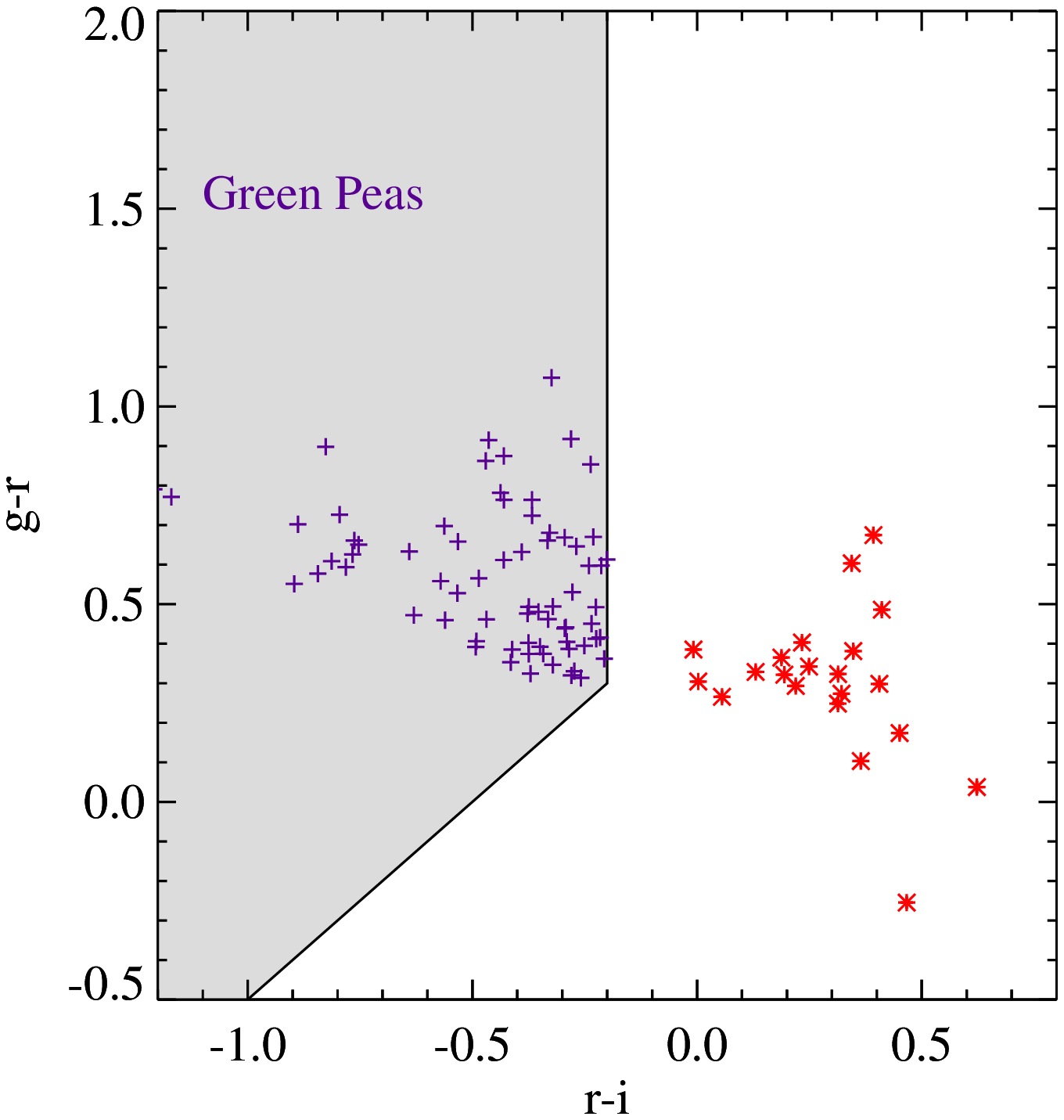}
\caption{The optical colours of our candidate sample (asterisks), compared to those of the optically-selected `Green Pea' population identified by Cardamone et al (2009, crosses). Photometric errors are typically smaller than the data points. Our candidates do not satisfy the green pea criteria. Although their spectra are generally blue (as required in the $u-r-z$ colour plane), they do not have the extreme excesses in the $r$-band identified by the $g-r-i$ colour selection.\label{fig:greenpeas}}
\end{figure}

In figure \ref{fig:greenpeas}, we compare the optical colours of our candidate sample with Green Peas presented by \citet{2009MNRAS.399.1191C}. Our sources do not satisfy the two-fold colour criteria required for identification as Green Peas. Although their spectra are generally blue (as required in the $u-r-z$ colour plane), they do not have the extreme excesses in the $r$-band identified by the $g-r-i$ colour selection.  This is not entirely surprising. The requirement for a very high equivalent width from emission lines in the $r$-band, but not in the $g$-band skews the distribution of emission line ratios in the Green Pea sample, and yields galaxies with a mean metallicity of 12-log(O/H)$\sim8.7$, rather higher than is typical for our sample, and comparable to Solar metallicity\footnote{12-log(O/H)$_\odot=8.69$ \citep{2001ApJ...556L..63A}}. It also biases the sample to very high star formation rates - with the required line widths giving a typical star formation rate of $\sim$30\,M$_\odot$\,yr$^{-1}$. Thus these sources are typically more metal rich and also forming stars at $\sim3-10$ times the rate of our targets. Given that star formation rate directly influences rest-UV flux, we would also expect the majority of Green Peas to fall outside our ultraviolet luminosity selection, were GALEX data for the whole sample to be available.

Green Peas are selected to be compact in SDSS data ($g$-band petrosian radius $<2.0\arcsec$, c.f. $<1.2\arcsec$ for our sample), although this requirement translates to a less strict constraint at the high redshift end of their sample. Nonetheless, there would likely be overlap between our sample and the Green Pea selection criteria were we to admit more UV-luminous sources. Those Green Peas which lie towards the low metallicity and lower redshift end of the Cardamone et al and similar selections may well be good analogues for luminous ($L_{UV}>L*_{z=6}$) Lyman break galaxies at high redshift in terms of star formation density. However, the $z\sim5$ LBA sample presented here remains distinct from the established Green Pea population and is more closely tuned to the properties of relatively faint galaxies currently being identified in deep surveys (see figure \ref{fig:lum-size}).

\section{Radio Continuum Follow-Up}\label{sec:radio}

Observations at radio wavelengths have the potential to further
elucidate the properties of galaxies such as these in two important
ways. They can reveal the presence of dust-extincted emission, i.e. a
more extended or intense emission source than that seen in the
ultraviolet bands where this sample was selected. Detection
of a significant excess of emission in radio wavelengths over
that predicted from the ultraviolet star formation rate could indicate that the ultraviolet sources are small regions of
intense star formation and low extinction embedded in a much larger and more evolved
galaxy \citep[][see also Stanway et al 2010, Davies et al
2012, 2013]{2010ApJ...710..979O} and so suggest that these sources would be more analoguous
to ULIRG-like super-starbursts than Lyman break galaxies. Alternately,
they may indicate the presence of an obscured active galactic nucleus
which could substantially effect the evolution and broadband
photometry of the system. These two effects are potentially
distinguishable based on the radio-frequency spectral slope of any
detected emission.

 \subsection{Observations}
We have undertaken an examination of the radio continuum flux at 3 and
6cm wavelengths in a subset of thirteen galaxies, randomly selected
from our larger sample of twenty-one targets.

We obtained continuum measurements of the selected galaxies over the
period 2012 Aug 27-29, using the Australia Telescope Compact Array
(ATCA) in its 6A configuration. The six antennae at the ATCA were
aligned along an East-West axis and the longest baselines were 6\,km
in length. We tuned the Compact Array Broadband Backend (CABB)
correlator such that one 2\,GHz IF was centered at 5500\,MHz (6\,cm)
and the second at 9000\,MHz (3\,cm), with full polarization
information collected simultaneously at both frequencies. Measurements
of PKS\,1934-638, the standard calibrator at the ATCA, were used for
absolute flux calibration. A bright, compact point source close to
each science target was used for atmospheric phase calibration.  Our
observations were associated with observing programme C2695 (PI:
Stanway).

Data were reduced using the dedicated software package {\sc miriad} \citep{1995ASPC...77..433S}
and radio frequency interference carefully flagged as a function of time on a
channel-by-channel basis. Each frequency band had 2048 one\,MHz-wide
spectral channels allowing interference to be restricted to a few
distinct channels. This is particularly important at 9000\,MHz where
powerful interference spikes occurred frequently in certain narrow
frequency ranges. Multifrequency synthesis images were constructed
from the 2\,GHz bandwidth at each frequency. In each image a number of
additional sources (often NVSS and occasionally 2MASS objects) were
detected. The images were cleaned using a Clark algorithm, constrained
to cut off at a level twice the noise standard deviation in the uncleaned
images.  Uniform
weighting was used to optimize suppression of
side-lobes in the imaging. All targets were expected to be point sources at 
the angular resolution of the ATCA.

Short tracks were taken across the full range of possible hour angles,
in order to optimize coverage of the $uv$-plane, with a total
on-source integration time of 2\,hours on each target. These sources are
relatively northern for the ATCA.  As a result, sources are below the
horizon for part of each potential twelve hour earth rotation
synthesis track, and the resulting synthesised beam is heavily
asymmetric. Typical beam sizes were $\sim2\arcsec$ in right ascension
and $\sim15-20\arcsec$ in declination at 5500\,MHz, with a beam position angle
close to zero. Flux uncertainty naturally scales with beam-size for a
point source and the noise is heavily correlated between adjacent
pixels due to uncertainties in the reconstruction of an
incompletely-sampled $uv$-plane.

The {\sc miriad} task `imfit' was used to measure the flux and noise
levels at the location of the galaxy, which was placed close to the
centre of the primary beam (8.5\,\arcmin\ FWHM at 5500\,MHz). Measured fluxes, beam sizes and
uncertainties for each target are given in table \ref{tab:radio}.

 \subsection{Results}
Eight sources are detected at better than $3\sigma$ significance in
our 5500\,MHz data. Of these, the brightest three sources were
independently detected in two or more observing sessions, each
separated by at least one day. Visual inspection suggests that the
weakest detection (in object 05083) is marginal and may or may not be
reliable, while a second (object 71294) is similarly faint and has a flux
measurement that may be affected by a brighter neighbour. All others
appear to be robust detections. None of the eight detected sources are
measurably resolved in the radio imaging.

The 9000\,MHz band is more heavily affected by noise than that at
5500\,MHz. Observations were taken simultaneously in the two bands,
and those at 9000\,MHz are naturally slightly shallower, although the
synthesised beam is smaller by a factor of two. The four brightest
sources at 5500\,MHz were also detected in the higher frequency
data. No other target was significantly detected at 9000\,MHz.  The
results of our radio observations for all targets are given in table
\ref{tab:radio}.

\begin{table*}
\begin{tabular}{lcccccccc}
Object ID & z & 5500\,MHz Flux & S/N & Beam size & 9000\, MHz & S/N & SFR & $\alpha$ \\ 
\hline\hline 
23734 & 0.108 & 220 $\pm$ 34 & 6.4 & 16\,$\times$\,2.0$\arcsec$ & 164 $\pm$ 31 & 5.2 & 5.5 $\pm$ 0.9 & -0.60 $\pm$ 0.15\\ 
10880 & 0.169 &  86 $\pm$ 24 & 3.6 & 18\,$\times$\,2.0$\arcsec$ & 43 $\pm$ 34 & 1.2 & 5.5 $\pm$ 1.5 & $<$-0.04\\ 
19220 & 0.188 &  60 $\pm$ 82 & 0.7 & 17\,$\times$\,2.0$\arcsec$ & 44 $\pm$ 32 & 1.4 & $<$13.3 & \\ 
54061 & 0.074 & 215 $\pm$ 38 & 5.6 & 19\,$\times$\,2.0$\arcsec$ & 136 $\pm$ 34 & 4.0 & 2.5 $\pm$ 0.4 & -0.93 $\pm$ 0.29\\ 
60392 & 0.120 & 54 $\pm$ 26 & 2.1 & 19\,$\times$\,1.9$\arcsec$ & 73 $\pm$ 41 & 1.8 & $<$1.6 & \\ 
27473 & 0.083 & 57 $\pm$ 23 & 2.5 & 16\,$\times$\,2.0$\arcsec$ & 47 $\pm$ 31 & 1.5 & $<$0.7 & \\ 
71294 & 0.146 & 72 $\pm$ 21 & 3.4 & 15\,$\times$\,2.0$\arcsec$ & 79 $\pm$ 53 & 1.5 & 3.4 $\pm$ 1.0 & $<$1.25\\ 
16911 & 0.097 & 91 $\pm$ 20 & 4.6 & 18\,$\times$\,2.0$\arcsec$ & 109 $\pm$ 27 & 4.1 & 1.8 $\pm$ 0.4 & -0.61 $\pm$ 0.12\\ 
10045 & 0.137 & 69 $\pm$ 26 & 2.6 & 15\,$\times$\,2.0$\arcsec$ & 112 $\pm$ 46 & 2.4 & $<$2.2 & \\ 
24784 & 0.113 & 151 $\pm$ 27 & 5.6 & 21\,$\times$\,1.9$\arcsec$ & 93 $\pm$ 43 & 2.2 & 4.2 $\pm$ 0.7 & $<$-0.70\\ 
62100 & 0.136 & 61 $\pm$ 25 & 2.4 & 20\,$\times$\,1.9$\arcsec$ & 67 $\pm$ 40 & 1.7 & $<$2.0 & \\ 
05083 & 0.146 & 76 $\pm$ 22 & 3.4 & 20\,$\times$\,1.9$\arcsec$ & 80 $\pm$ 32 & 2.5 & 3.6 $\pm$ 1.0 & $<$0.13\\ 
08755 & 0.164 & 159 $\pm$ 22 & 7.1 & 20\,$\times$\,1.9$\arcsec$ & 163 $\pm$ 33 & 5.0 & 9.6 $\pm$ 1.3 & 0.05 $\pm$ 0.01\\
\end{tabular}
\caption{Results from the radio observations at 5500 and 9000\,MHz, taken at the ATCA. Fluxes and 1\,$\sigma$ errors (measured for a point source at the location of the galaxy) are given in $\mu$Jy/beam. The beam size is given at 5500\,MHz and is half this size at 9000\,MHz. Measurements with an associated S/N below 3.0 are deemed non-detections. \label{tab:radio} The penultimate column gives the inferred star formation rate in solar masses per year, as described in section \ref{sec:radio}, with 2\,$\sigma$ limits where appropriate. For detected sources, the final column gives the observed-frame 5500MHz/9000MHz spectral index, $\alpha$, or limit thereupon (based on 2.5\,$\sigma$ non-detections at 9000\,MHz).}
\end{table*}

\subsection{Radio spectral slope}
Table \ref{tab:radio} also presents the radio spectral index for those
sources with robust detections in one or more wavebands. Where a
source is not detected at 9000\,MHz, we estimate a limit on the slope
based on a limit of 2.5 times the noise level for a point source (at
which point detection begins to be plausible in the images).

Radio continuum emission arises as a result either of AGN emission or
star formation. \tr{While synchrotron emission, arising from electrons
  accelerated in a strong magnetic field, is ubiquitous in radio
  sources of all types, the presence of other thermal or non-thermal
  processes can contribute different emission components, modifying
  the continuum spectral slope. For example, in ionised hydrogen
  clouds there is a substantial contribution from free-free or
  bremstrahlung radiation which arises when unbound electrons are
  scattered by the magnetic field of nearby ions \citep[see][ for a
    full explanation]{2002ira..book.....B}.} Hence, where AGN emission
dominates, a relatively steep spectral slope is expected,
theoretically becoming as steep as $\alpha\sim-2$ for bright quasars
with no star formation contribution \citep[][]{2002ira..book.....B},
and observed values steeper than $\alpha\sim-1$ \citep[e.g.][and
  references therein]{2012arXiv1201.3922G,2013ApJ...768...37C}.  By
contrast thermal free-free absorption and emission in star formation
regions are expected to flatten the spectrum (particularly at low and
high frequencies respectively) by contributing flux with
$\alpha\sim-0.1$ and the majority of star forming systems have a
fairly shallow slope ($\alpha\sim-0.5-0.6$)
\citep[see][]{1992ARA&A..30..575C,2003ApJ...588...99B}.  Therefore we
expect galaxies with ongoing star formation to have moderately
negative spectral slopes, and AGN or old stellar populations to be
steeper.

Measurements of the spectral index are possible for the four sources
detected in both bands of our observations (albeit with substantial
uncertainty). Of these, three galaxies have spectral slopes consistent
with $-\alpha\sim0.6-0.8$, as expected for faint star-forming sources
\citep[see ][]{2003ApJ...588...99B}. 

The above assumes that we are observing in a frequency regime
described by a single, unbroken power law - usually a safe
assumption. However aging of the synchrotron-emitting
(i.e. relativistic electron) population tends to steepen the radio
spectrum above a time-dependent critical value, since high energy
electrons decay more quickly than those at lower energies
\citep{1992ARA&A..30..575C}. This leads to a break in the spectrum
which increases in observed frequency with age. In all but the
youngest star forming galaxies this break will lie above the
frequencies considered here. One source in our sample (object 08755)
has a substantially flatter spectral slope than the others, with
$\alpha=0.05$. Spectral slopes this flat are relatively unusual, and
may indicate the presence of a spectral break at around
$\sim$6-8\,GHz, which would in turn suggest that the source of
non-thermal synchrotron emission in this source is very young
\citep{2012arXiv1201.3922G}.

None of the galaxies in this sample have spectral slopes steep enough
to preclude star formation dominated emission, but deeper
observations, and observations at more frequencies (both long-wards and
short-wards of 5500\,MHz) will be necessary to study the radio spectral
energy distribution of these sources in more detail. 

 \subsection{Star formation rates}

Where robustly detected, and with the possible exception of one
object, the radio spectral indices in this sample are consistent with
the flux arising from electrons accelerated by supernovae and their
remnants. Hence they should track the supernova (and so also star
formation) rate.

As a result, radio continuum flux can be straightforwardly converted
to an inferred star formation rate given an empirically determined
conversion factor. This factor is almost constant for stellar
populations forming stars at a constant rate and aged more than about
$10^8$\,years, but can vary for younger stellar populations \citep{2002A&A...392..377B}. 
We calculate the star formation rates inferred from our
continuum measurements at 5500\,MHz using the same prescription as
\citet{2003ApJ...588...99B} and \citet{2002ApJ...568...88Y}. Following
\citeauthor{2003ApJ...588...99B}, we set $\alpha=-0.6$, appropriate
for faint radio sources and consistent with those measured in our
sample.  We use $T_d=58$\,K and $\beta_\mathrm{FIR}=1.35$,
respectively the dust temperature and emissivity index \citep[again
  fixed to the values of][for comparison with previous
  studies]{2002ApJ...568...88Y}. At these wavelengths, the thermal
dust emission makes a negligible contribution, and setting $T_d=35$\,K
and $\beta_\mathrm{FIR}=2$ \citep[as found at $z\sim3$,][]{2013MNRAS.433.2588D} has no
measurable effect.

\begin{figure}
\includegraphics[width=\columnwidth]{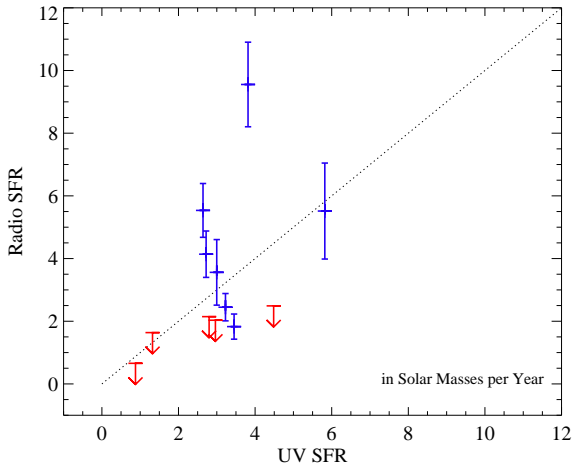}
\caption{Star formation rates inferred from the GALEX UV (as in figure
  \ref{fig:lum-size}) and ATCA 5500\,MHz continuum flux (described in
  section \ref{sec:radio}) for the subsample of our targets with ATCA
  imaging. 2\,$\sigma$ limits are shown where a source is
  undetected. Error bars show uncertainties due to photometric errors,
  but not systematic uncertainties in the star formation rate
  conversion factors. The rates are consistent to within a factor of a
  few both for detected sources and limits. Object 19220, for 
which relatively weak radio limits were obtained, is omitted from this figure. The star formation rates
  in these sources are similar to those typical in star forming
  galaxies at $z>5$ observed in deep fields. The non-detections at
  5500\,MHz may indicate young stellar populations, yet to establish a
  SNe-driven radio continuum in some cases, or that these sources have
  an unexpectedly steep radio spectral slope.\label{fig:radio_sfr}}.
\end{figure}

Figure \ref{fig:radio_sfr} illustrates the resultant radio-inferred
star formation rates (or limits thereupon) for our sample, in
comparison to the star formation rates inferred from their ultraviolet
flux\footnote{Using the same conversion factor applied in figure
  \ref{fig:lum-size}}. Of the sources with radio detections, the
inferred star formation rates derived using the two methods are
consistent within a factor of a few, with one source, object 08755,
showing a 4.3\,$\sigma$ excess in radio continuum, suggesting moderate
dust extinction may be reducing the observed ultraviolet flux. We note
that this source is an outlier in the sample in several ways: not only
does it show a radio excess, it also has a significantly flatter radio
spectral slope than those measured in other sources, and is also
amongst the reddest sources in the ultraviolet (with
$\beta\sim-1$). The sources without radio detections (5/13 targets)
are also broadly consistent with their ultraviolet-inferred star
formation rates at the 3\,$\sigma$ level, although in one case (object 19220), the
constraint is relatively weak and this source not considered further here.

\begin{figure}
\includegraphics[width=\columnwidth]{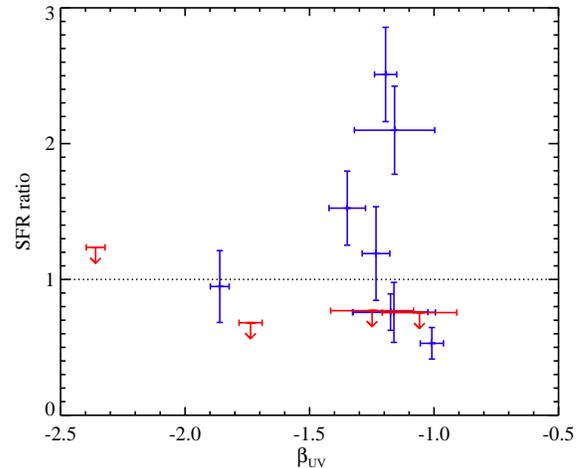}
\caption{The ratio of star formation rates derived from 5.5\,GHz flux to those derived from the ultraviolet
(i.e. a source with dusty star formation, or other sources of excess radio emission, would have a ratio
greater than one). Where a source is undetected in the radio, 2\,$\sigma$ limits are shown. Object 19220, for 
which relatively weak radio limits were obtained, is omitted from this figure. Four out of eight detections,
and three radio limits, are consistent with a deficit in radio flux relative to the ultraviolet. \label{fig:ratio}}
\end{figure}

However, both the detected fluxes and the limits on undetected objects hint that the
radio flux in the typical member of this population may actually be
somewhat deficient with respect to their ultraviolet flux. A point source
with a flux at least 2.5\,$\sigma$ times the background level should have been
detected in three of the four cases, and would also likely be accessible
to visual inspection, even if formally undetected. Similarly, of the sources that
were detected, half (4 out of 8) have measured radio fluxes below those predicted
from their UV luminosity (see figure \ref{fig:ratio}). This is a
tentative, but potentially interesting result if supported by future 
observations.

Given the uncertainties in the calibrations applied to derive star
formation rates here, it is useful to consider whether alternative
indicators might be useful. There is currently no deep near-infrared
data available for these sources (which will be surveyed by VISTA over
the next few years). As a result, SED fitting becomes ill-constrained
and will be explored in a later paper. Data from the WISE survey in
the mid-infrared does exist and, particularly in the 22$\mu$m W4 band,
can also used as a measure of star formation rate over a wide range in
luminosity.  However, without a deep $K$-band image to constrain the
location of red sources, deconvolution of blended sources in the
relatively shallow and low resolution mid-infrared images becomes a
substantial problem. As a result, non-detections and extended sources
in the WISE catalogues are unlikely to provide useful limits on our
targets.  Nonetheless, we identify four of our radio-observed sample
as relatively isolated and cleanly detected point sources in the
``ALLWISE'' data release of the W4-band all-sky
catalog\footnote{http://wise2.ipac.caltech.edu/docs/release/allwise/}.

\begin{figure}
\includegraphics[width=\columnwidth]{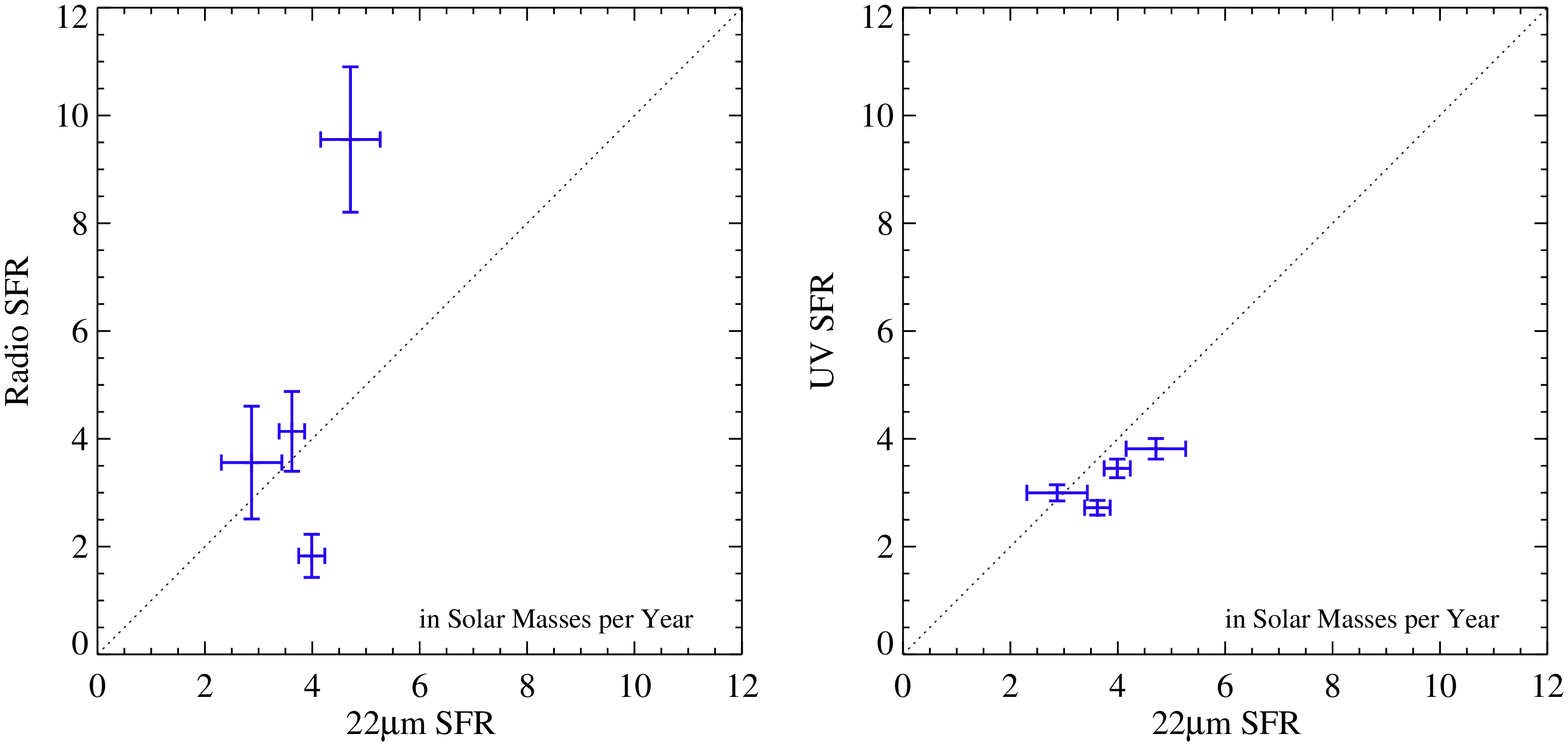}
\caption{Star formation rates inferred from the WISE 22$\mu$m band
  compared to their ultraviolet and 5500\,MHz continuum derived values
  for the four sources with good WISE detections. Error bars show
  uncertainties due to photometric errors, but not systematic
  uncertainties in the star formation rate conversion
  factors.\label{fig:wise_sfr}}
\end{figure}

Perhaps unsurprisingly, all four detected sources are also well
detected in the radio data. We provide a comparison between
ultraviolet-, 22$\mu$m- and radio-derived star formation rates for
these four sources in figure \ref{fig:wise_sfr}, using the star
formation rate conversion for the W4 band derived by
\citet{2012JApA...33..213S}. As the figure shows, the infrared-derived
star formation rates suggest that these sources contain relatively
little warm dust, exceeding the ultraviolet-derived values by only
$\sim$25\%. Interestingly, object 08755 (which has the highest star
formation rate in the sample and a flatter than expected radio
spectral slope, as discussed in the previous section), does not
present as an outlier in this respect, but is consistent with the
sample as a whole.

Given that the sources with strong radio detections might usually be
expected to be the most massive, and often the dustiest, of our
sample, the bulk of the non-detections and blended sources are likely
to prove similarly spare of warm dust and until data is available to
form a full SED fitting analysis, the ultraviolet-derived star
formation rates do not require a substantial dust correction.

As mentioned above, we convert radio flux to star formation using
an assumed radio continuum spectral slope $\alpha=-0.6$, consistent
with those we measure. Using a steeper spectral slope, $\alpha=-0.75$, results in
star formation rate estimates 25\% higher. However, even given an adjustment
of this magnitude, six of the sample still show inferred star formation
rates lower in the radio than the ultraviolet. Since the few sources bright and
isolated enough for WISE detections suggest that the ultraviolet star formation
rates might also increase by 25\%, the relative deficiency in radio flux 
remains.

A radio flux in excess of that predicted from the ultraviolet
photometry can be straightforwardly interpreted as indicating the
presence of dust-obscured star formation. A deficit in the radio
emission is more challenging to explain. Any star forming region from
which ultraviolet light can escape should present no impediment to the
escape of radio photons. Similarly, if the emission was influenced by
the presence of an AGN, the radio flux would be expected to exceed
that escaping in the ultraviolet. An intriguing possibility is that a
deficit in the radio may indicate a very young stellar population. The
ultraviolet flux in star forming systems arises from the photospheres
of the hottest, most massive stars, with a main sequence lifetime of a
few tens of Myr. By contrast, the radio flux is established by
supernovae and their remnants at the end of stellar lifetimes, and
thus takes longer to stabilize to the standard conversion factors
applied.

A young stellar population ($<$100\,Myr, approximately) would be
deficient in radio flux \citep[see][]{2002A&A...392..377B}. While this could
conceivably arise at the onset of continuous star formation, it could
also indicate a recent burst, or an exponentially rising star
formation rate with cosmic time in these sources \citep[as has been
suggested at high redshift by recent modeling and simulation,
e.g.][]{2010MNRAS.407..830M}.  Any scenario featuring a young, hot
stellar population is likely to lead to a bluer intrinsic ultraviolet
spectral slope, and thus runs counter to the tentative trend observed
in figure \ref{fig:ratio}, which is intriguing. The sources with the
bluest spectral slopes in fact show least evidence, one way or the
other, for a deficit in radio flux. Since emission from any regions of
moderately-dust extincted radio emission should, redden the
observed spectral slope, the relatively red ultraviolet colours of the
sources with the highest radio excesses are unsurprising. The red
colours of the sources with the largest radio \textit{deficits} are
rather more so and warrants more investigation.  Disentangling the
effect of stellar population age and reddening on the spectral slopes
in these galaxies will require a full analysis of the ultraviolet through
infrared spectral energy distribution of these sources and their
spectra and will be investigated in future work.

It is, of course, possible that the radio-deficient galaxies lie just
marginally below the detection level in our ATCA observations.
Further observations will be required to determine whether these
sources are indeed substantially lower in radio flux than expected. If
they are indeed young systems, this will strengthen their
interpretation as analogues for the generally young systems observed
at the highest redshift, and make them potentially useful models for
predicting the properties of distant galaxies at submillimetre/radio
wavelengths.

\section{Discussion and Conclusions}\label{sec:conc}

In this paper we have presented a new pilot sample of 21 local star forming
galaxies that present potential analogues for star forming galaxies in
the distant universe - that is, are potential Lyman break analogues
for the $z>5$ galaxy population.

We have demonstrated that these sources provide a good match to the
established or expected properties of the distant galaxy population,
in terms of star formation density, physical size, ultraviolet
luminosity and metallicity, and that they populate a region of these
parameter spaces either sparsely populated or unpopulated by existing
LBA samples. They show a weak trend towards blue ultraviolet colours
at low luminosities and low metallicities that mirrors those seen
in the high redshift galaxy population. 

Radio continuum investigations preclude the presence of a strong
obscured AGN in any of a subsample of 13 sources. Eight sources are
detected at a flux level consistent with that expected from their
ultraviolet luminosity, assuming that both radio and ultraviolet flux
arises from the same star forming population. The remaining sources
are undetected in the radio. The relatively low fraction of sources
with robust radio detections is intriguing, and, if supported by
future observations and forthcoming analysis of their spectral
energy distributions, may indicate that these sources are entirely
dominated by a very young stellar population.

LBAs such as this may prove extremely useful in interpreting the
limited data accessible through observations of faint and heavily
redshifted galaxies in the distant Universe. Deep surveys such as
CANDELS, and still more so the Hubble Ultra Deep Field campaign and Frontier Fields, are
probing a luminosity range, colour selection and star formation
density regime that isn't well explored by existing LBA samples. While
the number density of these very distant sources, and their
photometric colours in the rest-frame ultraviolet, are straightforward
to determine, very little detailed information can be extracted from their
photometry, and spectroscopy in most cases is limited to Lyman-alpha line emission and
weak detections of a low resolution spectral continuum. By contrast, LBAs can 
be investigated at a full range of wavelengths, from the ultraviolet through to
centimetre wavelengths, and detailed abundances and dust properties extracted
from optical through near-infrared photometry. 

A logical next step in our investigations is to explore the integrated
stellar populations in these galaxies - both through their optical
spectroscopy (which is inaccessible in high redshift galaxies) and
through analysis of their photometric spectral energy distribution
(which is directly comparable to the most commonly-applied technique
for analysis of high redshift samples). Forthcoming data from approved
LABOCA observations and the VISTA public surveys in the near-infrared,
forming a powerful combination with SDSS optical, GALEX ultraviolet
photometry and WISE (3-22\,$\mu$m) imaging should allow tight
constraints on dust content, stellar populations and mass to be
obtained through fitting of the spectral energy distributions.  We
also plan to explore their radio, millimetre and far-infrared
properties in more detail, now that radio detections have been secured
on an initial subsample, and are pursuing an approved programme of
integral field spectroscopy in the infrared using SINFONI on the ESO
VLT on a subset of objects. This information, will allow us to
determine the gas content, dust properties and physical conditions
within these young, compact systems, putting them in the wider context
of star forming galaxies at low redshifts, and comparing them to the
models commonly applied for high redshift star formation.

\section*{Acknowledgments}
LJMD acknowledges post-doctoral funding from the UK Science and Technology Facilities Council. 

This paper is based in part on data obtained at the Australia Telescope Compact Array associated with programme C2695. The Australia Telescope Compact Array is part of the Australia Telescope National Facility which is funded by the Commonwealth of Australia for operation as a National Facility managed by CSIRO.

It is also based in part on public data from the Sloan Digital Sky Survey DR7. Funding for the SDSS and SDSS-II has been provided by the Alfred P. Sloan Foundation, the Participating Institutions, the National Science Foundation, the U.S. Department of Energy, the National Aeronautics and Space Administration, the Japanese Monbukagakusho, the Max Planck Society, and the Higher Education Funding Council for England. The SDSS Web Site is http://www.sdss.org/. The SDSS is managed by the Astrophysical Research Consortium for the Participating Institutions. 

Also based in part on public data from GALEX GR6. The Galaxy Evolution Explorer (GALEX) satellite is a NASA mission led by the California Institute of Technology.


\bsp

\label{lastpage}


\begin{thebibliography}{99}
\bibitem[\protect\citeauthoryear{Abazajian et al.}{2009}]{2009ApJS..182..543A} 
Abazajian, K.~N., 
Adelman-McCarthy, J.~K., Ag{\"u}eros, M.~A., et al.\ 2009, ApJS, 182, 543 
\bibitem[\protect\citeauthoryear{Allende Prieto, Lambert, 
\& Asplund}{2001}]{2001ApJ...556L..63A} Allende Prieto C., Lambert D.~L., Asplund M., 2001, ApJ, 556, L63 
\bibitem[\protect\citeauthoryear{Baird}{1981}]{b1} Baird S.R., 1981,
ApJ, 245, 208
\bibitem[\protect\citeauthoryear{Berger et al.}{2003}]{2003ApJ...588...99B} 
Berger, E., Cowie, L.~L., Kulkarni, S.~R., et al.\ 2003, ApJ, 588, 99 
\bibitem[\protect\citeauthoryear{Bremer et al.}{2004}]{2004MNRAS.347L...7B} 
Bremer M.~N., Lehnert M.~D., Waddington I., Hardcastle M.~J., Boyce P.~J., 
Phillipps S., 2004, MNRAS, 347, L7 
\bibitem[\protect\citeauthoryear{Bouwens et al.}{2013}]{2013arXiv1306.2950B} 
Bouwens R.~J., et al., 2013, arXiv, arXiv:1306.2950
\bibitem[\protect\citeauthoryear{Bouwens et 
al.}{2012}]{2012ApJ...754...83B} Bouwens R.~J., et al., 2012, ApJ, 754, 83 
\bibitem[\protect\citeauthoryear{Bouwens et 
al.}{2007}]{2007ApJ...670..928B} Bouwens R.~J., Illingworth G.~D., Franx 
M., Ford H., 2007, ApJ, 670, 928 
\bibitem[\protect\citeauthoryear{Bressan, Silva, \& Granato}{2002}]{2002A&A...392..377B} 
Bressan A., Silva L., Granato G.~L., 2002, A\&A, 392, 377 
\bibitem[\protect\citeauthoryear{Calzetti et al.}{2000}]{2000ApJ...533..682C} 
Calzetti D., Armus L., Bohlin R.~C., 
Kinney A.~L., Koornneef J., Storchi-Bergmann T., 2000, ApJ, 533, 682 
\bibitem[\protect\citeauthoryear{Burgarella et 
al.}{2011}]{2011ApJ...734L..12B} Burgarella D., et al., 2011, ApJ, 734, L12 
\bibitem[\protect\citeauthoryear{Burke 
\& Graham-Smith}{2002}]{2002ira..book.....B} Burke B.~F., Graham-Smith F., 2002, ``An Introduction to Radio Astronomy: second edition'', Cambridge University Press, Cambridge, UK
\bibitem[\protect\citeauthoryear{Cardamone et 
al.}{2009}]{2009MNRAS.399.1191C} Cardamone C., et al., 2009, MNRAS, 399, 
1191 
\bibitem[\protect\citeauthoryear{Condon}{1992}]{1992ARA&A..30..575C} Condon J.~J., 1992, ARA\&A, 30, 575 
\bibitem[\protect\citeauthoryear{Condon et al.}{2013}]{2013ApJ...768...37C} 
Condon J.~J., Kellermann K.~I., Kimball A.~E., Ivezi{\'c} {\v Z}., Perley 
R.~A., 2013, ApJ, 768, 37 
\bibitem[\protect\citeauthoryear{Davies et al.}{2012}]{2012MNRAS.425..153D} 
Davies L.~J.~M., Bremer M.~N., Stanway E.~R., Mannering E., Lehnert M.~D., 
Omont A., 2012, MNRAS, 425, 153 
\bibitem[\protect\citeauthoryear{Davies et al.}{2013}]{2013MNRAS.433.2588D} 
Davies L.~J.~M., Bremer M.~N., Stanway E.~R., Lehnert M.~D., 2013, MNRAS, 
433, 2588 
\bibitem[\protect\citeauthoryear{Douglas et 
al.}{2010}]{2010MNRAS.409.1155D} Douglas L.~S., Bremer M.~N., Lehnert 
M.~D., Stanway E.~R., Milvang-Jensen B., 2010, MNRAS, 409, 1155 
\bibitem[\protect\citeauthoryear{Douglas et 
al.}{2009}]{2009MNRAS.400..561D} Douglas L.~S., Bremer M.~N., Stanway 
E.~R., Lehnert M.~D., Clowe D., 2009, MNRAS, 400, 561 
\bibitem[\protect\citeauthoryear{Douglas et 
al.}{2007}]{2007MNRAS.376.1393D} Douglas L.~S., Bremer M.~N., Stanway 
E.~R., Lehnert M.~D., 2007, MNRAS, 376, 1393 
\bibitem[\protect\citeauthoryear{Dunlop et al.}{2013}]{2013MNRAS.432.3520D} 
Dunlop J.~S., et al., 2013, MNRAS, 432, 3520 
\bibitem[\protect\citeauthoryear{Finkelstein et 
al.}{2012}]{2012ApJ...756..164F} Finkelstein S.~L., et al., 2012, ApJ, 756, 
164 
\bibitem[\protect\citeauthoryear{Georgakakis et 
al.}{2012}]{2012arXiv1201.3922G}  Georgakakis A., Grossi M., Afonso J., 
Hopkins A.~M., 2012, MNRAS, 421, 2223 
\bibitem[\protect\citeauthoryear{Grogin et al.}{2011}]{2011ApJS..197...35G} 
Grogin, N.~A., Kocevski, D.~D., Faber, S.~M., et al.\ 2011, ApJS, 197, 35 
\bibitem[\protect\citeauthoryear{Haberzettl et 
al.}{2012}]{2012ApJ...745...96H} Haberzettl L., Williger G., Lehnert M.~D., 
Nesvadba N., Davies L., 2012, ApJ, 745, 96 
\bibitem[\protect\citeauthoryear{Heckman et 
al.}{2005}]{2005ApJ...619L..35H} Heckman T.~M., et al., 2005, ApJ, 619, L35 
\bibitem[\protect\citeauthoryear{Heckman et 
al.}{2011}]{2011ApJ...730....5H} Heckman T.~M., et al., 2011, ApJ, 730, 5 
\bibitem[\protect\citeauthoryear{Heinis et al.}{2014}]{2014MNRAS.437.1268H} 
Heinis S., et al., 2014, MNRAS, 437, 1268 
\bibitem[\protect\citeauthoryear{Hoopes et al.}{2007}]{2007ApJS..173..441H} 
Hoopes C.~G., et al., 2007, ApJS, 173, 441 
\bibitem[\protect\citeauthoryear{Ibar et al.}{2009}]{2009MNRAS.397..281I} 
Ibar E., Ivison R.~J., Biggs A.~D., Lal D.~V., Best P.~N., Green D.~A., 
2009, MNRAS, 397, 281 
\bibitem[\protect\citeauthoryear{Izotov et 
al.}{2011}]{2011A&A...536L...7I} Izotov Y.~I., Guseva N.~G., Fricke K.~J., Henkel C., 2011, A\&A, 536, L7 
\bibitem[\protect\citeauthoryear{Jia et al.}{2011}]{2011ApJ...731...55J} 
Jia J., Ptak A., Heckman T.~M., Overzier R.~A., Hornschemeier A., LaMassa 
S.~M., 2011, ApJ, 731, 55 
\bibitem[\protect\citeauthoryear{Kewley \& Ellison}{2008}]{2008ApJ...681.1183K} 
Kewley L.~J., Ellison S.~L., 2008, ApJ, 681, 1183 
\bibitem[\protect\citeauthoryear{Madau, Pozzetti, \& Dickinson}{1998}]{1998ApJ...498..106M} 
Madau P., Pozzetti L., Dickinson M., 1998, ApJ, 498, 106 
\bibitem[\protect\citeauthoryear{Maiolino et 
al.}{2008}]{2008A&A...488..463M} Maiolino R., et al., 2008, A\&A, 488, 463 
\bibitem[\protect\citeauthoryear{Maraston et al.}{2010}]{2010MNRAS.407..830M} 
Maraston C., Pforr J., Renzini A., Daddi 
E., Dickinson M., Cimatti A., Tonini C., 2010, MNRAS, 407, 830 
\bibitem[\protect\citeauthoryear{Maraston}{2005}]{2005MNRAS.362..799M} 
Maraston C., 2005, MNRAS, 362, 799 
\bibitem[\protect\citeauthoryear{Nandra, Laird, \& Steidel}{2005}]{2005MNRAS.360L..39N} 
Nandra K., Laird E.~S., Steidel C.~C., 2005, MNRAS, 360, L39 
\bibitem[\protect\citeauthoryear{Oesch et al.}{2013}]{2013ApJ...772..136O} 
Oesch P.~A., et al., 2013, ApJ, 772, 136 
\bibitem[\protect\citeauthoryear{Oesch et al.}{2010}]{2010ApJ...709L..21O} 
Oesch P.~A., et al., 2010, ApJ, 709, L21 \bibitem[\protect\citeauthoryear{Overzier et 
al.}{2011}]{2011ApJ...726L...7O} Overzier R.~A., et al., 2011, ApJ, 726, L7 
\bibitem[\protect\citeauthoryear{Oh et al.}{2011}]{2011ApJS..195...13O} 
Oh K., Sarzi M., Schawinski K., Yi S.~K., 2011, ApJS, 195, 13 
\bibitem[\protect\citeauthoryear{Ono et al.}{2010}]{2010ApJ...724.1524O} 
Ono Y., Ouchi M., Shimasaku K., Dunlop J., Farrah D., McLure R., Okamura 
S., 2010, ApJ, 724, 1524 
\bibitem[\protect\citeauthoryear{Overzier et 
al.}{2010}]{2010ApJ...710..979O} Overzier R.~A., Heckman T.~M., 
Schiminovich D., Basu-Zych A., Gon{\c c}alves T., Martin D.~C., Rich R.~M., 
2010, ApJ, 710, 979 
\bibitem[\protect\citeauthoryear{Rafelski et 
al.}{2012}]{2012ApJ...755...89R} Rafelski M., Wolfe A.~M., Prochaska J.~X., 
Neeleman M., Mendez A.~J., 2012, ApJ, 755, 89 
\bibitem[\protect\citeauthoryear{Sault et al.}{1995}]{1995ASPC...77..433S} 
Sault, R.~J., Teuben, P.~J., 
\& Wright, M.~C.~H.\ 1995, Astronomical Data Analysis Software and Systems IV, 77, 433 
\bibitem[\protect\citeauthoryear{Shapley et al.}{2003}]{2003ApJ...588...65S} 
Shapley A.~E., Steidel C.~C., Pettini M., Adelberger K.~L., 2003, ApJ, 588, 65
\bibitem[\protect\citeauthoryear{Shi et al.}{2012}]{2012JApA...33..213S} 
Shi F., Kong X., Wicker J., Chen Y., Gong Z.-Q., Fan D.-X., 2012, JApA, 33, 
213 
\bibitem[\protect\citeauthoryear{Stanway et 
al.}{2010}]{2010MNRAS.407L..94S} Stanway E.~R., Bremer M.~N., Davies 
L.~J.~M., Lehnert M.~D., 2010, MNRAS, 407, L94 
\bibitem[\protect\citeauthoryear{Stanway, McMahon, 
\& Bunker}{2005}]{2005MNRAS.359.1184S} Stanway E.~R., McMahon R.~G., Bunker A.~J., 2005, MNRAS, 359, 1184 
\bibitem[\protect\citeauthoryear{Stanway, Bunker, 
\& McMahon}{2003}]{2003MNRAS.342..439S} Stanway E.~R., Bunker A.~J., McMahon R.~G., 2003, MNRAS, 342, 439 
\bibitem[\protect\citeauthoryear{Steidel et 
al.}{2004}]{2004ApJ...604..534S} Steidel C.~C., Shapley A.~E., Pettini M., 
Adelberger K.~L., Erb D.~K., Reddy N.~A., Hunt M.~P., 2004, ApJ, 604, 534 
\bibitem[\protect\citeauthoryear{Tanvir et al.}{2012}]{2012ApJ...754...46T} 
Tanvir N.~R., et al., 2012, ApJ, 754, 46 
\bibitem[\protect\citeauthoryear{Th{\"o}ne et 
al.}{2013}]{2013MNRAS.428.3590T} Th{\"o}ne C.~C., et al., 2013, MNRAS, 428, 
3590 
\bibitem[\protect\citeauthoryear{Verma et al.}{2007}]{2007MNRAS.377.1024V} 
Verma A., Lehnert M.~D., F{\"o}rster Schreiber N.~M., Bremer M.~N., Douglas 
L., 2007, MNRAS, 377, 1024 
\bibitem[\protect\citeauthoryear{Wilkins et 
al.}{2013}]{2013MNRAS.430.2885W} Wilkins S.~M., Bunker A., Coulton W., 
Croft R., Matteo T.~D., Khandai N., Feng Y., 2013, MNRAS, 430, 2885 
\bibitem[\protect\citeauthoryear{Wilkins et al.}{2011}]{2011MNRAS.417..717W} 
Wilkins S.~M., Bunker A.~J., Stanway E., 
Lorenzoni S., Caruana J., 2011, MNRAS, 417, 717 
\bibitem[\protect\citeauthoryear{Yun 
\& Carilli}{2002}]{2002ApJ...568...88Y} Yun M.~S., Carilli C.~L., 2002, ApJ, 568, 88 
\end{thebibliography}
\end{document}